\documentclass[twocolumn,showpacs,preprintnumbers,amsmath,amssymb]{revtex4}

\usepackage{fancyheadings}

\usepackage{amsmath}
\usepackage{amsfonts}
\usepackage{amssymb}
\usepackage{graphicx}

\begin{document}

\title{Branched Quantization}
\author{Alfred Shapere$^1$ and Frank Wilczek$^2$\vspace*{.1in}}
\affiliation{$^1$Department of Physics and Astronomy,
University of Kentucky, Lexington, Kentucky 40502 USA\\
$^2$Center for Theoretical Physics,
Department of Physics, Massachusetts Institute of Technology,
Cambridge, Massachusetts 02139 USA}

\vspace*{.3in}
\begin{abstract}
We propose a method for quantization of Lagrangians for which the Hamiltonian, as a function of momentum, is a branched function with cusps.   Appropriate boundary conditions, which we identify, insure unitary time evolution. In special cases a dual (canonical) transformation maps the problem into a problem of quantum mechanics on singular spatial manifolds, which we also develop.  Several possible applications are indicated.   
\end{abstract}

\maketitle

\thispagestyle{fancy}

\bigskip


Quantum mechanics, like classical mechanics -- and unlike, say, the standard model of fundamental interactions -- is more a conceptual framework than a concrete account of specific physical phenomena.    Physically interesting models based on quantum mechanics bring in additional structure.  The analytical versions of classical mechanics have been a fruitful source of inspiration in this regard.   There are procedures that allow one to pass from a wide variety of classical Hamiltonian systems to quantum models.   In those constructions, guiding principles include Bohr's correspondence principle, which basically asks that the quantum model should reproduce, approximately, the original classical dynamics in appropriate limits; maintenance of symmetry; and internal consistency -- specifically, unitary time evolution.    There are also important but less fully developed, and possibly less rigorous, procedures using path integrals that allow one to pass directly from classical Lagrangian systems, including some singular ones, to quantum models \cite{henneaux}.    In any case, we should regard the construction of quantum models as a creative process, open to innovation.    

Recently, in constructing models of possible time crystals \cite{classicalTXTal}, \cite{quantumTXTal}, we were led to consider Lagrangians involving higher than quadratic powers of the time derivatives, and specifically the deceptively simple
\begin{equation}\label{timeXTalL}
L ~=~ \frac{1}{4} {\dot x}^4 - \frac{\kappa}{2} {\dot x}^2
\end{equation}
In the interesting case $\kappa > 0$, the Hamiltonian for this Lagrangian is singular.   Since the momentum involves a cubic in velocity,
\begin{equation}\label{timeXTalP}
p ~=~ {\dot x}^3 - \kappa {\dot x}
\end{equation}
we can have either one or three real values of $\dot x$ corresponding to a given value of $p$, depending on whether  $| \dot x | -  \sqrt{\frac{\kappa}{3}}$ is positive or negative.  Thus the energy function
\begin{equation}\label{timeXTalE}
E ~=~ \frac{\partial L}{\partial {\dot x}} \dot x - L ~=~ \frac{3}{4} {\dot x}^4 - \frac{\kappa}{2} {\dot x}^2 \, ,
\end{equation}
expressed in terms of $p$, is a multivalued function with cusps.   See Figure 1.

\begin{figure}[ht]
\includegraphics[scale=0.6]{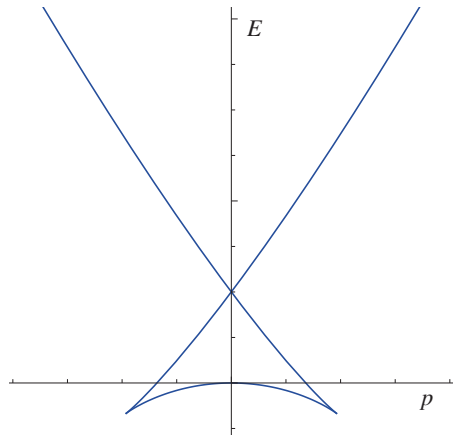}
\caption{Energy as a function of momentum.}
\end{figure}   

Here we propose methods for constructing quantum models corresponding to broad classes of classical systems with branched structures in either momentum or position space.   

{\it Branched quantization.}
Because the Hamiltonian based on Eqns.\,(\ref{timeXTalP}, \ref{timeXTalE}) is not a single-valued function of $p$, and yet energy must surely qualify as an observable, $p$ does not supply a complete set of commuting observables.  Therefore it will not be sufficient to label states with wave functions in (conventional) momentum space.    On the other hand $E$ is a single-valued function of $\dot x$, so we might expect to construct wave functions $\psi (\dot x)$.   As $\dot x$ runs monotonically from $-\infty \rightarrow \infty$, $p(\dot x)$ evolves along the trajectory indicated in Figure 1, reversing direction at the cusps.   This suggests that we consider wave functions that depend on $p$ locally, but accommodate the back-tracking.   Thus, denoting by $p_\pm \equiv \pm \sqrt{\frac{\kappa}{3}}$ the cusp points, we have three components to the wave function, namely $\psi_1(p)$ for $-\infty < p \leqslant p_+$, $\psi_2(p)$ for $p_- \leqslant p \leqslant p_+$, and $\psi_3(p)$ for $p_- \leqslant p < \infty$.   Note that all three components cover the range $p_- \leqslant p \leqslant p_+$.    

A crucial issue is how the different branches join together, i.e., what are appropriate boundary conditions.    In this connection it is instructive to consider a more general class of Lagrangians than Eqn.\,(\ref{timeXTalL}), bringing in a quadratic potential $V(x) = \frac1{2} {\alpha}x^2$.    Directly from the Schr\"odinger equation we have an equation for the probability density $\rho_\mu (p, t) \equiv  \psi_\mu(p,t)^* \psi_\mu(p, t)$, defined on the branch labeled by $\mu$ (where $\mu=1,2,3$ for $p_-< p < p_+$, $\mu=1$ for $p<p_-$, and $\mu = 3$ for $p> p_+$):
\begin{equation}\label{probabilityCurrent}
\frac{\partial \rho_\mu}{\partial t} = i \left(\psi_\mu^* H \psi_\mu - (H^* \psi_\mu^*) \psi_\mu \right) = -\frac{i\alpha}{2} \bigl( \psi_\mu^* \frac{\partial^2 \psi_\mu}{\partial p^2} - \frac{\partial^2 \psi_\mu^*}{\partial p^2} \psi_\mu \bigr)
\end{equation}
using, in the second step, $V(x) \rightarrow V(i\frac{\partial}{\partial p})$.  This substitution implements the basic Heisenberg commutation relation, and also reflects the role of $p$ as the generator of spatial translations.   From Eqn.\,(\ref{probabilityCurrent}) we infer an equation of the current-conservation type for $\rho\equiv \sum \rho_\mu$
\begin{equation}\label{currentConservation}
\frac{\partial \rho}{\partial t} + \frac{\partial j}{\partial p} ~=~ 0 \, ; \ \ \ 
j ~\equiv~\sum_\mu \frac{i\alpha}{2} \bigl( \psi_\mu^* \frac{\partial \psi_\mu}{\partial p} - \frac{\partial\psi_\mu^*}{\partial p} \psi_
\mu \bigr) \end{equation}
Eqn.\,(\ref{probabilityCurrent}) will lead to conservation of the integrated probability $\int \rho$ if we can drop contributions from $j$ at the endpoints.  (Note that $j$ receives contributions from two branches at each endpoint $p_\pm$.)   We also require our boundary conditions to be linear, so that our Hilbert space will support superposition, and so that they lead to a physically sensible eigenvalue problem for $H$.   The choices
\begin{equation}\label{boundaryConditions}
\psi_1 (p_+) ~=~ \psi_2 (p_+) \, ; \ \ \ \ \ 
\frac{\partial \psi_1}{\partial p} (p_+) ~=~ - \frac{\partial \psi_2}{\partial p} (p_+) 
\end{equation}
and their analogues at $p_-$ manifestly give the required cancellation in $j$.   Now consider the eigenvalue problem for the time-independent Schr\"odinger equation.   For $\alpha \neq 0$, we get a second-order differential equation for $\psi(p)$.   Thus, on each branch, for each value of energy, there are two disposable constants, making six altogether.   Eqn.\,(\ref{boundaryConditions}) gives us four constraints among these constants, and normalizability (absence of growing modes) at $p \rightarrow \pm \infty$ gives us two more (one for each of $\psi_1, \psi_3$).    Thus the number of constraints matches the number of constants, as in conventional quantum potential theory.  

If $\alpha =0$ the derivative conditions in Eqn.\,(\ref{boundaryConditions}) are not required and should not be imposed.  On the other hand if $V$ is a higher-order polynomial, we must require augmented boundary conditions.   We will discuss those presently, after introducing a different (dual) viewpoint.

{\it Dual Viewpoint.}
One can hardly fail to notice that the manipulations we performed in momentum space, in connection with probability conservation, resemble manipulations usually performed in position space.   Thus it is natural to consider what our models look like after the substitution $p \rightarrow x$, $x \rightarrow -p$, which preserves the structure of quantum mechanics.  After this substitution, our multi-valued kinetic energy becomes something perhaps less unconventional, that is, a multivalued potential.    Indeed, we may think of a wire with kinks, as in Figure 2a.   Intermediate values of $x$ are triply represented, and physical conditions will be different at different points {\it along the wire}, even if they are represented by the same $x$, so a branched wave function is manifestly appropriate to describe this physical system.   

\vspace{0.1in}
\begin{figure}[ht]
\includegraphics[scale=1]{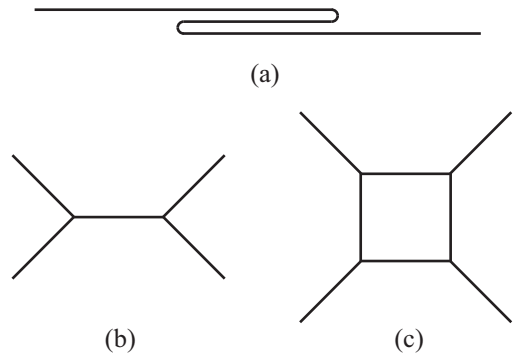}
\caption{(a) A wire with kinks.  (b) A wire network with two junctions.  (c) Box graph for a network with four junctions and a loop.}
\end{figure}   

From this dual point of view our quadratic potential $V(x) \rightarrow V(p) = \frac12 {\alpha}p^2$ becomes the conventional kinetic energy of a particle with mass $m = 1/\alpha$, and the branched kinetic term becomes a multivalued potential $W(x)$ in position space.   Thus we have wave functions $\psi_1 (x)$ defined for $-\infty <  x \leqslant x_+$ subject to $W_1(x)$, $\psi_2 (x)$ defined for $x_- \leqslant  x \leqslant x_+$ subject to $W_2(x)$, and $\psi_3 (x)$ defined for $x_-  \leqslant x < \infty$ subject to $W_3(x)$, and boundary conditions similar to Eqn.\,(\ref{boundaryConditions}), after the obvious substitutions of $x$ for $p$.   

Now let us consider a quartic potential $V(x) = x^4 + \alpha x^3 + \beta x^2 + \gamma x$.  In dual variables this leads to a kinetic energy that is a quartic polynomial in $p$, $H=p^4 - \alpha p^3 + \beta p^2 - \gamma p+W(x)$.   We find a probability current (in the dual $x$ space)
\begin{eqnarray}\label{quarticCurrent}
j ~&=&~ \frac{1}{i} \bigl( \psi^\dagger \frac{\partial^3 \psi}{\partial x^3} - \frac{\partial \psi^\dagger}{\partial x} \frac{\partial^2 \psi}{\partial x^2}  + \frac{\partial^2 \psi^\dagger}{\partial x^2} \frac{\partial \psi}{\partial x} - \frac{\partial^3 \psi^\dagger}{\partial x^3} \psi \bigr) 
 \nonumber \\
~&&~ - \alpha ( \psi^\dagger \frac{\partial^2 \psi}{\partial x^2} - \frac{\partial \psi^\dagger}{\partial x} \frac{\partial  \psi}{\partial x }  + \frac{\partial^2 \psi^\dagger}{\partial x^2}  \psi ) \nonumber\\ 
~&&~+ i \beta  ( \psi^\dagger \frac{\partial \psi}{\partial x} - \frac{\partial\psi^\dagger}{\partial x} \psi ) 
+ \gamma \, \psi^\dagger \psi
\end{eqnarray}
where $\psi$ is a column vector with the requisite number of components $\psi_\mu$ in each momentum range. 
We will insure conservation of probability with the boundary conditions
\begin{equation}\label{quarticBC}
\frac{\partial^n \psi_1}{\partial x^n} ~=~ (-1)^n \frac{\partial^n \psi_2}{\partial x^n} \ \ \ \ \ \ \ 0 \leq n \leq 3
\end{equation}
at the kinks, provided that $\alpha$ and $\gamma$ change sign between the branches.  This augmentation of the boundary conditions also leads to a good eigenvalue problem, since we have both twice as many disposable constants and twice as many conditions as in the quadratic case.   

Inspired by the wire analogy it is natural, and for later purposes instructive, to consider networks analogous to the geometries of electric circuit theory, where we put quantum dynamics on graphs \cite{kuchment}.   Let us consider what is required to insure no flow of probability into a node where several lines indexed by $\mu$ come together.  If the momentum dependence on each line is simply $p^2$, and we orient each line so all coordinates $x^\mu$ flow into the node, then the ``Kirchoff" boundary conditions
\begin{eqnarray}\label{simpleBC}
\psi_1 ~&=&~ \psi_2 ~=~ ...  \\
0 ~&=&~ \sum\limits_\mu \frac{\partial \psi_\mu}{\partial x^\mu}
\end{eqnarray}
insure that no probability accumulates at the node.   These natural conditions give good eigenvalue problems for the ``Compton'' tree graph and the box graph displayed in Figure 2b,c (and many others).  In the Compton graph, Eqn.\,(\ref{simpleBC}) yields $2\times 3 = 6$ conditions at the two nodes, which together with four conditions at infinity for the external legs gives 10 conditions, as is appropriate to five lines with two disposable constants each.   In the box graph, Eqn.\,(\ref{simpleBC}) yields $4\times 3 = 12$ conditions at the four nodes, which together with four conditions at infinity gives 16 conditions, as is appropriate for eight lines with two disposable constants each.  Though it is simple and natural, this is by no means the only possible set-up consistent with the general requirements of the framework of quantum mechanics \cite{kuchment}.   We can vary both the Hamiltonians and the boundary conditions.   

To illustrate the possibilities, consider that each line might have its own kinetic Hamiltonian $\alpha_\mu p^2 + \beta_\mu p$ (and of course its own potential).  We can insure conservation of probability with boundary conditions of the form
\begin{equation}
\psi_\mu  ~=~ \kappa_\mu \psi 
\end{equation}
at the node, together with 
\begin{equation}
\sum\limits_\mu \alpha_\mu \kappa_\mu^* \frac{\partial \psi_\mu}{\partial x^\mu} ~=~ 0
\end{equation}
so long as  $\sum\limits_\mu \beta_\mu | \kappa_\mu |^2 = 0$.  With quartics, the possibilities proliferate.

{\it Unfolding}.
Quadratic models with branching in the dual viewpoint can be unfolded, upon which the eigenvalue problem assumes a conventional form.   In equations: If we write define a real variable $\chi$ with
\begin{eqnarray}
\chi ~&\equiv&~ \ \ \ \, x - x_+ +  x_-\ \ \ {\rm for} \ \ \chi \leqslant x_- \nonumber \\
\chi ~&\equiv&~ -x + x_+ + x_- \ \ \  {\rm for} \ \ x_- \leqslant \chi \leqslant x_+ \nonumber \\
\chi ~&\equiv&~ \ \ \ \, x + x_+ - x_- \ \ \ {\rm for} \ \ x_+ \leqslant \chi
\end{eqnarray}
then as $\chi$ evolves monotonically from $-\infty \rightarrow \infty$ it covers each branch of $x$ uniquely (as $\dot x$ covered $p$ in our earlier discussion).  
The boundary conditions Eqn.\,(\ref{boundaryConditions}), with $p \rightarrow x$ become, upon transcription into the unfolding variable $\chi$, the statement that $\psi(\chi)$ and its derivative are continuous.    Similarly, the boundary conditions Eqn.\,(\ref{quarticBC}) for quartics unfold into continuity for $\psi(\chi)$ and its first three derivatives.    

{\it Potentials}.
Now we return to our original problem, the issue of quantizing the kinetic Lagrangian of Eqn.\,(\ref{timeXTalL}) allowing for a general potential $V(x)$.  Inspired by the preceding unfolding procedure, we formulate our wave function in term of a variable that locally reduces to $\pm p$ plus a c-number, but covers all three branches following the same flow directions as $\dot \phi$.   Thus we introduce
\begin{eqnarray}
\xi ~&\equiv&~ \ \ \ \, p - p_+ +  p_-\ \ \ {\rm for} \ \ \xi \leqslant p_- \nonumber \\
\xi ~&\equiv&~ -p + p_+ + p_- \ \ \  {\rm for} \ \ p_- \leqslant \xi \leqslant p_+ \nonumber \\
\xi ~& \equiv&~ \ \ \ \, p + p_+ - p_- \ \ \ {\rm for} \ \ p_+ \leqslant \xi
\end{eqnarray}
and the decomposition of wave functions 
\begin{eqnarray}
\psi (\xi) ~&=&~ \psi  (\xi) \bigl(1 - H(\xi - p_-)\bigr)\nonumber\\
&&~~+~ \psi (\xi ) \bigl(H(\xi - p_-)-H(\xi - p_+)\bigr)  \nonumber\\
&&~~+~\psi(\xi) H(\xi - p_+) \nonumber\\
&\equiv&~ \psi_1(\xi) + \psi_2(\xi) + \psi_3(\xi)\, ,
\end{eqnarray}
where $H$ is the Heaviside function.  In this formulation $p$ is realized (piecewise) as a modified multiplication operator, with slightly different modifications on each branch.  We can use that fact to write the operator $V$ as an explicit kernel in $\xi$ space.   Thus we transform $\psi_1 (\xi)$ to $x$ space, where $V$ acts as multiplication, and transform back as follows:
\begin{eqnarray}\label{actionOfV}
u(x) &\equiv& \int e^{ipx} \psi_1 (\xi^\prime)\,  \frac{d\xi^\prime}{2\pi} = \int e^{i (\xi^\prime + p_+ - p_-)x} \psi_1 (\xi^\prime) \, \frac{d\xi^\prime}{2\pi} \nonumber \\
({\hat V} u) (x) &=& V(x) u(x) \nonumber \\
({\hat V} \psi) (\xi) &=& \int e^{-ipx} ({\hat V} u) (x) dx = \int\limits^{p_-}_{-\infty} K_1(\xi - \xi^\prime) \psi (\xi^\prime) \, \frac{d\xi^\prime}{2\pi} \nonumber \\
&&K_1(\xi - \xi^\prime) = \int e^{-i(\xi - \xi^\prime)x} V(x)  \, dx 
\end{eqnarray}
Note that the result of $\hat V$ acting on $\psi_1$ generally does not vanish for $\xi > p_-$.   It is realized as an operator of the Wiener-Hopf type.   

Performing the same manipulations on in the other intervals, we arrive at  $K_2 = K_1^*, K_3 = K_1$ and 
\begin{eqnarray}\label{trialVRealization}
({\hat V \psi}) (\xi) &=& \int \Bigl( K_1 (\xi - \xi^\prime) \psi_1(\xi^\prime) + K_2 (\xi - \xi^\prime) \psi_2(\xi^\prime) \nonumber\\ 
&&+ K_3 (\xi - \xi^\prime) \psi_3(\xi^\prime) \Bigr) \, d\xi^\prime
\end{eqnarray}
The peculiarity of $K_2$ arises from the reversed flow of $p$, as a function of $\xi$, in the medial interval.   
In the symmetric case $V(x) = V(-x)$ all the $K$s are real and equal, and $\hat V$ becomes an ordinary convolution operator.   

If $V(x)$ is not symmetric, however, we must reconsider our procedure, because the ${\hat V}$ defined in Eqn.\,(\ref{trialVRealization}) is not Hermitean.   Indeed, although each $K_j$ satisfies the hermiticity condition $K_j(\xi^\prime, \xi) = K_j^* (\xi, \xi^\prime)$  the full kernel 
\begin{eqnarray}
K(\xi^\prime, \xi) &=& K_1 (\xi^\prime, \xi) (1- H(\xi - p_-) ) \nonumber\\
&&+~ K_2 (\xi^\prime, \xi) (H(\xi - p_-) - H(\xi - p_+)) \nonumber\\
&&+~ K_3 (\xi^\prime, \xi) H(\xi - p_+)
\end{eqnarray}
does not.   Thus to reach a consistent quantization we must impose $K_2 = K_1$ also (and not $K_2 = K_1^*$).   The sign changes  for $\alpha$ and $\gamma$ required in Eqn.\,(\ref{quarticBC}) foreshadowed this conclusion.   By adopting this modified quantization condition, we lose the Heaviside functions and arrive at a (manifestly Hermitean) convolution.  The modified quantization condition entails that the basic commutation relation $ [\xi, x ] = -i $ involves the unfolded $\xi$, not the mechanical $p$.   

For long-range potentials $V(x)$ the formal definition of $\hat V$ by Fourier transformation leads to derivatives of $\delta$ functions, which must be defined through integration by parts on the momentum-space wave functions.   In this way we make contact with our earlier discussion of polynomial potentials, and see why smoothness conditions connecting the different zones in $\psi$ are required, that become more demanding as the order of $V$ increases.


The eigenvalue problem for $H$, written in the $\xi$ representation, generally leads to an integral equation, rather than the boundary value problem for a differential equation, because $H$ is not polynomial in momenta.  This can be solved directly.   Alternatively, energy eigenstates $\psi_E$ with energy $E$ are characterized by the conditions
\begin{eqnarray}
\langle \psi_E | [ H, {\cal O} ] | \psi_E \rangle &=& 0 \nonumber \\
\langle \psi_E | H | \psi_E \rangle &=& E \nonumber \\
\langle \psi_E | \psi_E \rangle &=& 1
\end{eqnarray}
for a complete basis of operators ${\cal O}$, or by the variance conditions
\begin{eqnarray}
\langle \psi_E | H^2  | \psi_E \rangle \ \langle \psi_E | {\cal O}  {\cal O}^\dagger  | \psi_E \rangle &+& \langle \psi_E | H {\cal O}  {\cal O}^\dagger H| \psi_E \rangle =  \nonumber \\   \langle \psi_E | H   | \psi_E \rangle &\times& \langle \psi_E | \{ H ,  {\cal O}  {\cal O}^\dagger \} | \psi_E \rangle   \nonumber \\
\langle \psi_E | H | \psi_E \rangle &=& E \nonumber \\
\langle \psi_E | \psi_E \rangle &=& 1
\end{eqnarray}
We can solve these approximately by iterative (Newton) root-finding and minimization methods, respectively, as explained in detail in \cite{newtonMethod}.  The required evaluations of expectation values, given our explicit expressions, provide a practical approach to the eigenvalue problem.   

Eigenvalue problems for several examples of branched quantization with different potentials $V(x)$ are analyzed in \cite{examples}. 

{\it Another unfolding method.} Another natural choice of an unfolding coordinate for $p$ is $\dot x$.   
The phase space coordinates $(x,\dot x)$ are noncanonical, but one can formulate a symplectic structure and a Poisson bracket for them  \cite{examples}\cite{zhao}, which reduces to the standard Poisson bracket in the momentum range for which the map between  
$\dot x$ and $p$ is invertible:
\begin{equation}
\{ F,G\} ~=~ \frac{1}{3\dot x^2 -\kappa} \left[ \frac{\partial F}{\partial x}\frac{\partial G}{\partial \dot x}
- \frac{\partial F}{\partial \dot x}\frac{\partial G}{\partial x}\right]
\end{equation}
With this Poisson bracket and the Hamiltonian 
\begin{equation}
H(x,\dot x) ~=~ \frac{3}{4} {\dot x}^4 - \frac{\kappa}{2} {\dot x}^2 + V(x) \, ,
\end{equation}
one obtains Hamilton's equation $\dot F = \{ F,H \}$, which
reproduces the equation of motion derived from Eqn.~(\ref{timeXTalL}).  
This Hamiltonian formulation demonstrates that time evolution is well-defined (except for when $\dot x = \pm\sqrt{\kappa/3}$, where the symplectic structure degenerates and nondeterministic motion may occur \cite{classicalTXTal}) and offers an alternative approach to quantization \cite{examples}.

\vspace{0.1in}
{\it Comments}:
\begin{enumerate}
\item  A very common and fruitful procedure in analyzing quantum many-body problems, is to model the effect of interactions on a given particle by an effective one-body Hamiltonian (or Lagrangian), solving the one-body problem, and constructing a many-body wave function as a suitable product, {\it e. g}. a Slater determinant.    By widening the class of candidate one-body Hamiltonians, we can hope to extend this sort of analysis to wider classes of systems.   ``Swallowtail'' structures similar to Figure 1 have appeared in the description of Bose-Einstein condensates in lattice traps \cite{swallowtail}.  Dynamical mean field theory \cite{dmft} generates complicated time dependence in a time-translation invariant energy functional, as a consequence of interactions.   Polynomial truncation of such time dependence, using substitutions of the type $\bigl(x(t) - x(t-\delta)\bigr)^n  \to \delta^n {\dot x}^n$ \cite{classicalTXTal}, with retention of spatial structure, brings us to the sort of models considered here.    

\item  A particularly interesting case arises for periodic potentials $V(x)$.   In that case, famously, conventional kinetic terms lead to band structures: The energy becomes a multivalued function of the quasi-momentum.     Our branched Hamiltonian already for $V=0$ has a sort of band structure, associated with the branches, in a region of momentum $p_- < p  < p_+$ where the limiting $p_-, p_+$ are determined by the form of the kinetic energy, not by any spatial periodicity.   With a periodic potential added, both sources of banding are effective.   Especially interesting is the possibility of describing dynamically induced insulating behavior (Mott phenomenon) at filling fractions determined dynamically by the value of $\kappa$.

\item Our earlier work on time crystals was somewhat schizophrenic.   In the classical case \cite{classicalTXTal}, we found systems with motion in their ground state using kinetic Lagrangians of the type considered above.   In the quantum case \cite{quantumTXTal}, not knowing how to treat such Lagrangians, we used a different mechanism, based on a more conventional kinetic term, that depended on the discreteness of (generalized) quantum angular momentum.   A possible experimental realization has been proposed \cite{timeXTalExpt}. With the method here described we can quantize the classical time crystals, and  construct much more general candidate models of quantum time crystals.

\item We initially passed to the dual models to guide our intuition, but they appear to have considerable independent interest.   The central observation, that appropriate, fairly simple boundary conditions, both on wave functions and (odd) interactions, appears to allow construction of unitary quantum mechanics on a wide variety of singular manifolds.  One can even relax the boundary conditions, at the cost of allowing probability to flow into and out of the designated points -- in other words, by allowing the points to have internal degrees of freedom.   This procedure appears natural, specifically, in the modeling of black holes, where (in the Euclidean formalism) the horizon appears as a sphere attached to a point.    

\end{enumerate}


{\it Acknowledgements}:  AS is supported in part by NSF Grants PHY-0970069 and PHY-0855614.  FW is supported in part by DOE grant DE-FG02-05ER41360.

\end{document}